\documentclass{emulateapj}

\usepackage{amsmath}
\usepackage{epstopdf}
\usepackage{natbib}
\usepackage{color}
\usepackage{hyperref}
\newcommand{\be}{\begin{equation}}
\newcommand{\ee}{\end{equation}}

\newcommand{\tmin}{$\Delta t_{\rm min\; }$}
\newcommand{\tninty}{$T_{\rm 90}$}

\newcommand{\fermi}{\textit{Fermi }}
\newcommand{\swift}{\textit{Swift }}

\slugcomment{Accepted to ApJ}

\shorttitle{GRB Minimum Timescales}
\shortauthors{Golkhou, Butler \& Littlejohns}

\def\lessim{\mathrel{\hbox{\rlap{\hbox{\lower4pt\hbox{$\sim$}}}\hbox{$<$}}}}
\def\gtrsim{\mathrel{\hbox{\rlap{\hbox{\lower4pt\hbox{$\sim$}}}\hbox{$>$}}}}

\begin{document}

\setlength{\pdfpageheight}{\paperheight}
\setlength{\pdfpagewidth}{\paperwidth}

\title{The Energy-Dependence of GRB Minimum Variability Timescales}

\author{V. Zach Golkhou\altaffilmark{1,2}
, Nathaniel R. Butler\altaffilmark{1,2}
\& Owen M.  Littlejohns\altaffilmark{1,2}}

\altaffiltext{1}{School of Earth and Space Exploration,
Arizona State University, Tempe, AZ 85287, USA}
\altaffiltext{2}{Cosmology Initiative,
Arizona State University, Tempe, AZ 85287, USA}

\begin{abstract}
We constrain the minimum variability timescales for 938 GRBs observed by the \textit{Fermi}/GBM instrument prior to July 11, 2012.  The tightest constraints on progenitor radii derived from these timescales are obtained from light curves in the hardest energy channel.  In the softer bands -- or from measurements of the same GRBs in the hard X-rays from \swift -- we show that variability timescales tend to be a factor 2--3 longer.  Applying a survival analysis to account for detections and upper limits, we find median minimum timescale in the rest frame for long-duration and short-duration GRBs of 45 ms and 10 ms, respectively.  Fewer than 10\% of GRBs show evidence for variability on timescales below 2 ms.  These shortest timescales require Lorentz factors $\gtrsim 400$ and imply typical emission radii $R \approx 1 {\times} 10^{14}$ cm for long-duration GRBs and $R \approx 3 {\times} 10^{13}$ cm for short-duration GRBs.  We discuss implications for the GRB fireball model and investigate whether GRB minimum timescales evolve with cosmic time.
\end{abstract}


\maketitle

\section{Introduction}
\label{sec:intro}

Gamma-Ray Bursts (GRBs) are the most luminous explosions in the Universe,
originating at cosmological distances and releasing ${\sim} 10^{\rm 51}$ ergs over timescales
of seconds to tens of seconds.
The gargantuan energy release is accompanied by a very rapid and stochastic temporal variability
in the gamma-ray emission.
The \swift \citep{2004ApJ...611.1005G} 
 and \fermi Space Telescopes \citep{2009ApJ...702..791M}
 have deepened immensely our understanding of these cosmological beacons \citep[e.g.,][]{2013FrPhy...8..661G}.

The pulses observed in prompt GRB light curves often have a
Fast Rising Exponential Decay (FRED) profile \citep{1996ApJ...459..393N}. 
The time profiles can have a broad morphological
diversity in both the number of and duration of these pulses.
In the external shock model for GRBs, shells of material
produced by the GRB impact material in the circumburst medium \citep[e.g.,][]{1992MNRAS.258P..41R}.
Unless the circumburst medium is highly-clumped \citep{1999ApJ...512..683F}, this process
tends to produce a smooth GRB light curve in contrast
to the rapid temporal variability observed in many GRBs.
Under  the  internal  shock mechanism  \citep{1994ApJ...430L..93R},  a
variable  central  engine  emits  a  relativistic  outflow  comprised of
multiple shells with different  Lorentz factors, $\Gamma$. As faster
shells  collide with  slower shells, kinetic  energy  is converted  to
radiation, and multiple shell collisions can lead to a complex GRB light curve \citep[e.g.,][]{1994ApJ...430L..93R}.

Traditional   duration  measures  such  as   $T_{\rm  90}$
\citep{1993ApJ...413L.101K}, which describes the time during which the
central 90\%  of prompt gamma-ray  counts are received,  only describe
bulk  emission  properties of  the  burst.  Such a  duration
does not capture  information concerning individual  collisions between
shells.   Instead, detailed temporal  analyses that  probe variability
over a function of timescales are required.\par

A variety of time series analyses have previously been used to explore
the  rich  properties  of  prompt  GRB light  curves.   These  include
structure              function              (SF)             analyses
\citep{1994ApJ...433..494T,1994MNRAS.268..305H,1996A&A...306..395C,1997MNRAS.286..271A},
autocorrelation            function           (ACF)           analyses
\citep{1993ApJ...408L..81L,1995ApJ...448L.101F,1996ApJ...464..622I,2004A&A...418..487B,2012ApJ...749..191C},
and     Fourier    power     spectral    density     (PSD)    analyses
\citep{2000ApJ...535..158B,2001ApJ...557L..85C,2010ApJ...722..520A,2012MNRAS.422.1785G,2013MNRAS.431.3608D}. Compared
to power-spectral analyses,  the SF approach is less  dependent on the
time         sampling         \citep{1999ASPC..159..293P}.         In,
\citet{2014ApJ...787...90G},  Paper  I  hereafter,  we  developed  and
applied  a fast  (i.e.   linear)  and robust  SF  estimator, based  on
non-decimated  Haar  wavelets,  to  measure  the  minimum  variability
timescale, $\Delta t_{\rm min}$, of \swift GRBs.  We used the first-order SF of light
curves as  measured by the \textit{Swift} Burst  Alert Telescope (BAT;
\citealt{2005SSRv..120..143B})  to  infer  the shortest  timescale  at
which a GRB exhibit \textit{uncorrelated temporal variability}.\par

One limitation of  the work  presented  in Paper I is  that we  only
consider the  variability timescale  using light curves measured over
the narrow 15--350 keV energy band of \textit{Swift}/BAT.
A fixed and narrow energy band in the observer frame
would probe different regions of the intrinsic GRB spectra, because
GRBs are known to occur  over a wide range of  redshifts (see
e.g. \citealt{2009Natur.461.1258S,2009Natur.461.1254T,2011ApJ...736....7C,2012ApJ...752...62J}).
Previous studies have shown that GRB pulses 
vary in duration as a function of energy, 
with harder energy channels having a lower observed duration
\citep{1995ApJ...448L.101F,1996ApJ...459..393N}.
Working at higher energies -- where pulses are narrower -- also has the potential to provide tighter limits
on variability timescales.

We wish to use the broad \textit{Fermi}  Gamma-ray  Burst Monitor  (GBM;\citealt{2009ApJ...702..791M})
energy coverage to overcome this limitation and to effectively standardize a measure of the minimum variability timescale
by studying the energy evolution and/or evaluating the minimum timescale in
a fixed rest frame bandpass.
Broad energy coverage can potentially also allow us to disentangle
the role the ejecta velocity plays in relating radius to minimum timescale and to understand
how minimum timescales measured for different instruments should be compared \citep[see, e.g.,][]{2014arXiv1408.3042S}.
Also, it is important to note that the GBM provides very fine time resolution ($2 \mu$s) event mode data
for the full GRB and not just the first 1--2 s as was the case for \textit{BATSE} \citep[e.g.,][]{2000ApJ...537..264W}.

In the discussion below, we begin with a brief application and summary of the method outlined
in detail in Paper I.  We then
investigate  how  \tmin  depends  on  energy for  a  large  sample  of
\textit{Fermi}/GBM GRBs (Section \ref{sec:edep}).  We compare \tmin estimates
from \swift and \fermi for bursts detected in common to demonstrate stability and accuracy of error estimates (Section \ref{sec:joint}).
We then use spectral hardness to standardize the \tmin estimate (Section \ref{sec:hardness}) and
conclude by deriving constraints on the sample Lorentz factors and emission radii (Section \ref{sec:prog})
and by investigating potential evolution of \tmin with cosmic time (Section \ref{sec:cosmic_time}).

\section{Data}
\label{sec:data}

We   consider   949 GRBs   published   in  the   second
\textit{Fermi}/GBM  GRB   catalog  \citep{2014ApJS..211...13V},
spanning
the first four years of  the \textit{Fermi} mission (between 2008 July
14$^{\rm th}$ and 2012 July  11$^{\rm th}$, inclusive).
Event  lists  for  942  of  these  bursts  were downloaded from  the  online
\textit{Fermi}/GBM                                                burst
catalog\footnote{\url{http://heasarc.gsfc.nasa.gov/W3Browse/fermi/fermigbrst.html}}.\par

We analyze the  \textit{Fermi}/GBM Time-Tagged Event (TTE) data for each of the 12 sodium iodide
scintillators.  We only consider
those detectors in which each  GRB was brightest, as listed in column 2 of Table
7 in  \citet{2014ApJS..211...13V}.  Typically, this entails using
event lists for three detectors for each GRB.  Following \citet{2013MNRAS.432..857M}, we extract
200~$\mu$s binned light curves in the full (8 keV -- 1 MeV)
energy range. We  also extract light curves in  four energy channels
of an  equal logarithmic width  (8--26, 26--89, 89--299  and 299--1000
keV). These channels are referred to as channels 1, 2, 3, and 4 below.

To  remove background  counts from  the \textit{Fermi}/GBM  we  employ
a two-pass procedure.   Using the estimates of $T_{\rm 90}$ from 
Table 7  of \citet{2014ApJS..211...13V}, we bin
each light  curve at a resolution of $T_{\rm 90}/100$
and fit a linear
background model.  The background is initially determined considering two regions of each light curve, both $T_{\rm 90}$ in
length, occurring immediately before  and after the identified period of
burst emission.  Using the  background subtracted light curve, we then
estimated $T_{\rm 100}$ by accumulating a further 5\% of the $T_{\rm 90}$ interval counts
outward from both the  beginning and end of $T_{\rm 90}$. The
second pass at fitting a linear background is then conducted, masking
out  all  bins included  in  the  total  $T_{\rm 100}$ region.   This  second
background fit is then scaled to subtract the
predicted background counts in the fine-time-resolution light curve.
Our analysis -- which identifies variations on timescales short
compared to the overall burst durations -- does not require the fitting of background models more complex than linear.

We analyze the background-subtracted burst counts in the full $T_{\rm 100}$ region following the procedure outlined in Paper I.
One change is made to the algorithm to optimize for the detection of signal variations on short timescales: instead of
re-binning the 200 $\mu$s light curve to a fixed S/N per bin, we weight the unbinned light curve by the denoised \citep[following,][]{1997ApJ...483..340K}
signal.  This zeros-out portions of the light curve containing no signal and permits use of the full $T_{\rm 100}$ region without
adversely affecting our ability to identify variations on much shorter timescales.

For 109 bursts in  the second \textit{Fermi}/GBM  GRB catalog
which   also   have    \textit{Swift}   high-energy   prompt   coverage,
the \textit{Swift}/BAT   data  were   obtained  from   the  \textit{Swift}
Archive\footnote{\url{ftp://legacy.gsfc.nasa.gov/swift/data}}.          Using
calibration files from the  2008-12-17 BAT data release, we construct
100~$\mu$s  light curves, in  the full  15--350~keV BAT  energy range.
We use the standard  \textit{Swift}
software  tools:  {\sc   bateconvert},  {\sc  batmaskwtevt}  and  {\sc
  batbinevt}.  Further   details  regarding  the   extraction  of  the
\textit{Swift}/BAT light curves can  be found in Paper I.

\section{Discussion and Results}
\label{sec:DisRes}

In Paper I, we demonstrate the power of a novel, wavelet-based method -- the Haar-Structure Function (denoted $\sigma_{X,\Delta t}$) -- to robustly extract the shortest variability timescale of GRBs detected by \textit{Swift}/BAT.  In this work, we implement our technique on GRBs detected by the \textit{Fermi}/GBM instrument, which is sensitive to a much broader range of energies.
We obtain constraints on the minimum variability timescales for 938 of 949 GRBs reported in the second \textit{Fermi}/GBM GRB catalog \citep{2014ApJS..211...13V}.
Of these, we are able to confirm the presence of a linear rise phase (see Section \ref{sec:edep}) in the Haar-Structure Function on short timescales for 528 GRBs.  We quote upper-limit values for the remainder.
Most (421) of the bursts in this sub-sample are long-duration ($T_{\rm 90}>3$ s) GRBs.  In this sub-sample, 24 GRBs  have measured redshift, $z$.
The temporal specifications of all 938 GRBs discussed here are determined using fully-automatic software and are presented in Table 2. \par

\subsection{Studying the Energy-Dependence of \tmin}
\label{sec:edep}

It has been recognized for decades \citep[e.g.,][]{1995ApJ...448L.101F,1996ApJ...459..393N} that a defining feature
of GRB emission is a narrowing of pulse profiles observed in increasingly higher energy bands.
As a result, durations measured by different instruments can be different \citep[e.g.,][]{2012MNRAS.424.2821V}.
Durations also appear to depend on redshift, perhaps as a result of the dependence on bandpass:
recently, \citet{2013ApJ...778L..11Z} have found evidence that $T_{\rm 90}$ duration -- when $z$ is known and used to evaluate the GRB duration in a fixed rest frame energy band -- may correlate linearly with redshift as is expected from cosmological time dilation.  This result is quite sensitive to the particular choice of binning in the analysis \citep[see,][]{2014MNRAS.444.3948L}.
Here, we seek to understand whether our measure of shortest duration in GRBs is also highly-dependent upon the observed energy band, and on the instrument detecting the GRB, in particular.

\begin{figure}[!ht]
\begin{center}
\includegraphics[width=.45\textwidth,height=5.8in]{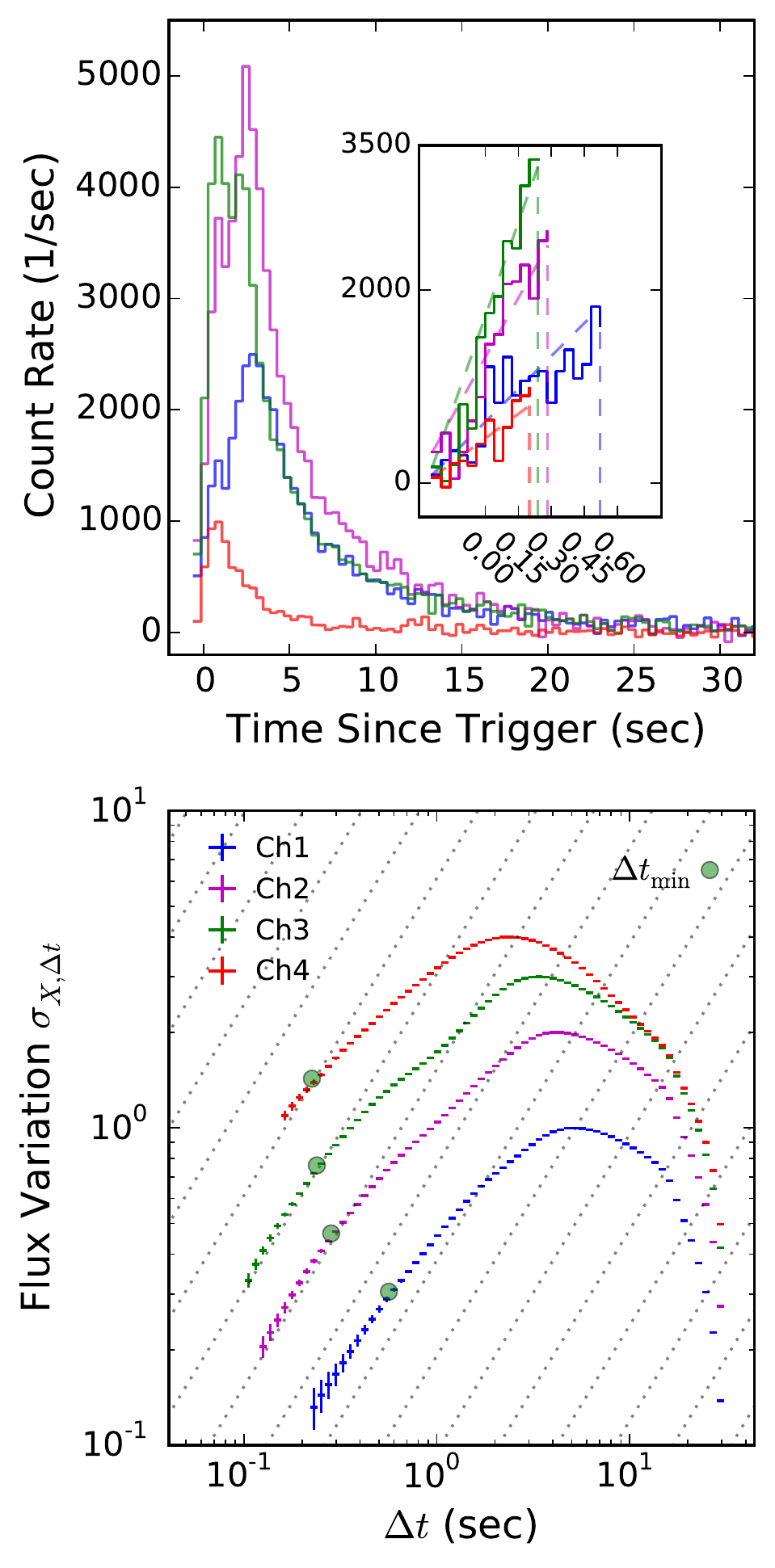} 
\caption{\small 
Top panel:  \textit{Fermi}/GBM light curves of the GRB~110721A split in 4 different energy bands.
Bottom panel: The Haar wavelet scaleogram $\sigma_{X,\Delta t}$, rescaled for plotting purposes, corresponding to each channel versus timescale $\Delta t$ for GRB~110721A.
We derive minimum timescales (marked with green circles) -- $0.56 \pm 0.09$ s,  $0.28 \pm 0.05$ s, $0.24 \pm 0.04$ s,
and $0.22 \pm 0.04$ s for the channels 1, 2, 3, and 4, respectively -- which increase in lower energy bands. 
In the top panel, the inset displays the the pulse rise with finer time binning, with dashed lines dropped onto the x-axis to
demark the derived $\Delta t_{\rm min}$ values for each channel.
}
\label{fig:sample}
\end{center}
\end{figure}

The prompt GBM Gamma-ray light curve for GRB~110721A, split in 4 energy bands, and our derived $\sigma_{X,\Delta t}$ curve for each channel are shown in Figure \ref{fig:sample}.  There is a clear evolution in \tmin with bandpass, decreasing from the softest to the hardest energy band. 
To guide the eye, several lines of constant $\sigma_{X,\Delta t} \propto \Delta t$ are also plotted.   The expected Poisson level (i.e., measurement error) has been subtracted away, leaving only the flux variation expected for each channel.

Briefly, we review here how our $\Delta t_{\rm min}$ is identified.
A general feature observed in our GRB scaleograms, provided there is sufficient signal-to-noise
ratio (S/N), is a linear rise phase relative to the Poisson noise.   
Poisson noise sets a floor on the shortest measurable timescale (denoted $\Delta t_{\rm S/N}$, with $\Delta t_{\rm S/N} \approx 0.1$ s for channels 2 and 3 in Figure \ref{fig:sample}, bottom).
Unlike previous studies by other authors \citep{2013arXiv1307.7618B,2013MNRAS.432..857M,2000ApJ...537..264W}, we do not implicate the shortest {\it observable} timescale as $\Delta t_{\rm min}$.  Instead, we recognize that pulses can be temporally smooth on short timescales.
The departure from this smoothness creates a break in the scaleogram, and this in turn defines our timescale $\Delta t_{\rm min}$ for temporally un-smooth variability.
Naturally, this timescale also corresponds to a length-scale, which must be reconciled with GRB progenitor models (Section \ref{sec:prog}).

We now focus on the softest energy band of GRB~110721A, denoting the light curve as $X(t)$.
Although there is excess signal present on timescales as short as $\Delta t=0.4$ s (Figure \ref{fig:sample} - channel 1), these timescales correspond to a region of the plot where the first order SF rises linearly timescale, $\sigma_{X,\Delta t} \equiv \left\langle |X(t+\Delta t)-X(t)|^2 \right\rangle_t^{1/2} \propto \Delta t$.  
(Here, $\left\langle . \right\rangle_t$ denotes an average over time $t$.)
We interpret this linear rise as an indication that the GRB exhibits temporally-smooth variations on these timescales (i.e., $X(t+\Delta t) \approx X(t) + X'(t) \Delta t$), while changing to exhibit temporally-unsmooth variations on longer timescales.  The $\sigma_{X,\Delta t}$ points deviate significantly from the $\sigma_{X,\Delta t} \propto \Delta t$ curve at $\Delta t_{\rm min} = 0.56 \pm 0.09$ s.  
This is the timescale of interest, describing the minimum variability time for uncorrelated variations in the GRB.  This timescale is associated with the initial rise of the GRB in this channel, as can be seen from the Figure \ref{fig:sample} inset.

The value for $\Delta t_{\rm min}$ is found by fitting a broken powerlaw to the $\sigma_{X,\Delta t} $ data points below the peak, assuming that $\sigma_{X,\Delta t}$ initially rises linearly with $\Delta t$ (see, also, Paper I) until flattening at $\Delta t_{\rm min}$.  Uncertainties quoted here and below for $\Delta t_{\rm min}$ are determined by direct propagation of errors and correspond to 1$\sigma$ confidence.  If the lower-limit on $\Delta t_{\rm min}$ falls below the lowest measurable timescale (i.e., $\Delta t_{\rm S/N}$), we report only the 1$\sigma$ upper limit for $\Delta t_{\rm min}$.

For this particular burst, \tmin evolves from the hardest energy band to the softest energy band as one might expect:
the softest energy band of a burst has longer minimum variability timescale compared to the hardest energy band of that burst.
On timescales longer than $\Delta t_{\rm min}$, $\sigma_{X,\Delta t}$ is flatter than $\sigma_{X,\Delta t} \propto \Delta t$, indicating the presence of temporally-variable structure on these timescales.  On a timescale of about 6 s, $\sigma_{X,\Delta t}$ begins turning over as we reach the timescales (tens of seconds) describing the overall emission envelope.  We are not concerned here with those longer timescale structures, although we do note that $\sigma_{X,\Delta t}$ provides a rich, aggregate description of this temporal activity.

\begin{figure}[ht!]
\centering
\includegraphics[width=.47\textwidth]{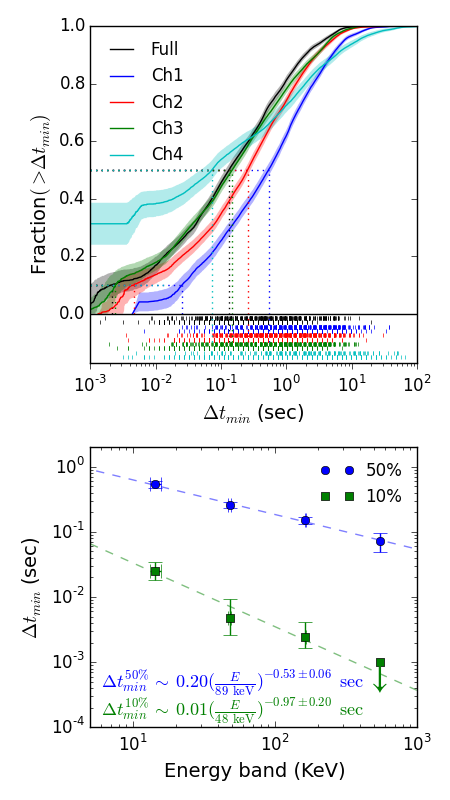} 
\caption{\small 
Top panel: The KM cumulative estimation curve of all long-duration GRBs in \fermi sample for each energy band including shaded 1$\sigma$ region around each bandpass. The dotted lines show the $50^{\rm th}$ percentile and the lowest $10^{\rm th}$ percentile for each bandpass. 
The location of the of measured $\Delta t_{\rm min}$ values (top tics) and upper-limits (bottom tics of the same color) are shown in the sub-panel.
Bottom panel: The KM median estimation of $\Delta t_{min}$ versus energy band and the lowest $10^{\rm th}$ percentile of \tmin values versus energy band, including error bars. 
Note: since the KM cumulative estimation curve of channel 4 does not cross the 10\% line, we plot an upper-limit.   
}
\label{fig:km}
\end{figure}

In order to characterize and measure the average \tmin for the \fermi sample as a function of spectral energy band, we utilize the Kaplan-Meier (KM; \citealt{kaplan58}, see also \citealt{1985ApJ...293..192F}) survival analysis.  This is necessary because many bursts only permit upper limit measurements of $\Delta t_{\rm min}$.
Figure \ref{fig:km} summarizes how the minimum variability timescale varies with energy band. The KM cumulative plots -- including the shaded 1$\sigma$ error region -- for each bandpass and the full (all channels combined) \textit{Fermi}/GBM energy range are shown in the top panel. The sample $50^{\rm th}$ percentiles (i.e., medians) and the lowest $10^{\rm th}$ percentiles (shown with the dotted-lines in the top panel of Figure \ref{fig:km}) are plotted in the bottom panel. 
Table 1 summarizes the corresponding values.  Since the KM cumulative estimation curve of channel 4 does not cross the 10\% limit line, there is no value reported in Table 1 for this case. 
The reported values clearly show the tendency of increasing \tmin with decreasing energy band.  Because we tend to find a clear association between \tmin and the rise time of the shortest GRB pulse (also, Paper I), this confirms that GRB pulse structures are narrower at higher energy and that understanding this effect is important for understanding any implications drawn from $\Delta t_{\rm min}$.

The KM median values of \tmin versus energy band are well-fitted by a line $\Delta t_{min}^{50\%} = 0.20 (E /89 \rm \, keV)^{-0.53 \pm 0.06}$ s (with reduced $\chi^2 = 0.64$). The derived power-law index here is in agreement with the power-law index of the relationship found for the average pulse width of peaks as a function of energy (\citealt{1995ApJ...448L.101F} and also from \citealt{1996ApJ...459..393N}).
The KM estimation of the lowest 10\% of \tmin values versus the energy band can also be fitted by a power-law, with a steeper index, $\Delta t_{min}^{10\%} = 0.01 (E / \rm 48 \, keV)^{-0.97 \pm 0.20}$ s (with reduced $\chi^2 = 1.4$). 
The steeper index indicates that rare GRBs, which tend to be bright and spectrally hard GRBs, allow for tighter constraints on minimum timescales. This shifts the typical minimum timescales to smaller values as compared to those found for the bulk of the population.  We explore the minimum timescale dependence on S/N and spectral hardness below for individual GRBs.

\begin{deluxetable}{lcccc}
\tablecolumns{5}
\tablewidth{0pc}
\tabletypesize{\scriptsize}
\tablecaption{The Kaplan-Meier median and $10^{\rm th}$ percentile timescales for
long-duration GRBs.\label{tab:tab1}}
\tablehead{
\colhead{Band} & \colhead{$\Delta t_{\rm min}^{\rm 50\%}$} & \colhead{$\Delta t_{\rm min}^{\rm 10\%}$} & \colhead{Number} & \colhead{Number} \\
\colhead{(keV)} & \colhead{(msec)} & \colhead{(msec)} & \colhead{Detected} & \colhead{Upper Limit}
}
\startdata
8--26 & $540 \pm 67$ & $25 \pm 8$ & 395 & 307\\[4pt]
26--89 & $260\pm 26$ & $4_{-2}^{+4}$ & 431 & 319\\[4pt]
89--299 & $150_{-18}^{+23}$ & $2_{-1}^{+2}$ & 413 & 335\\[4pt]
299--1000 & $72_{-21}^{+24}$ & ... & 156 & 278\\[4pt]
8--1000 & $130 \pm 18$ & $2_{-1}^{+5}$ & 421 & 334
\enddata
\end{deluxetable}

\begin{figure*}[!ht]
\begin{minipage}{.96\textwidth}
  \includegraphics[trim = 0mm 8mm 0mm 0mm,width=1.\linewidth]{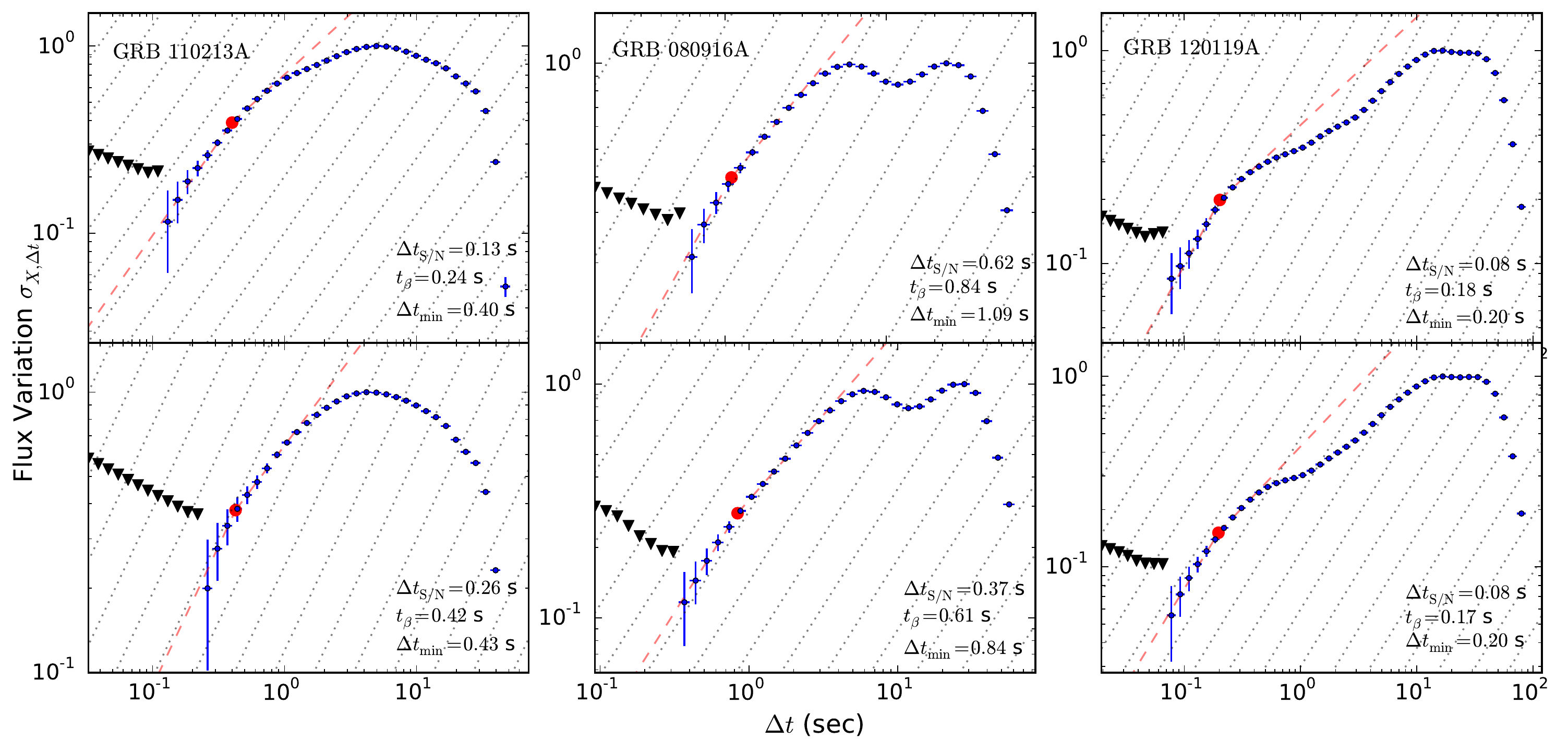}
  \label{fig:sclgm}
\end{minipage}
\begin{minipage}{.95\textwidth}
  \includegraphics[trim = 4mm 7mm 0mm 0mm,width=1.\linewidth]{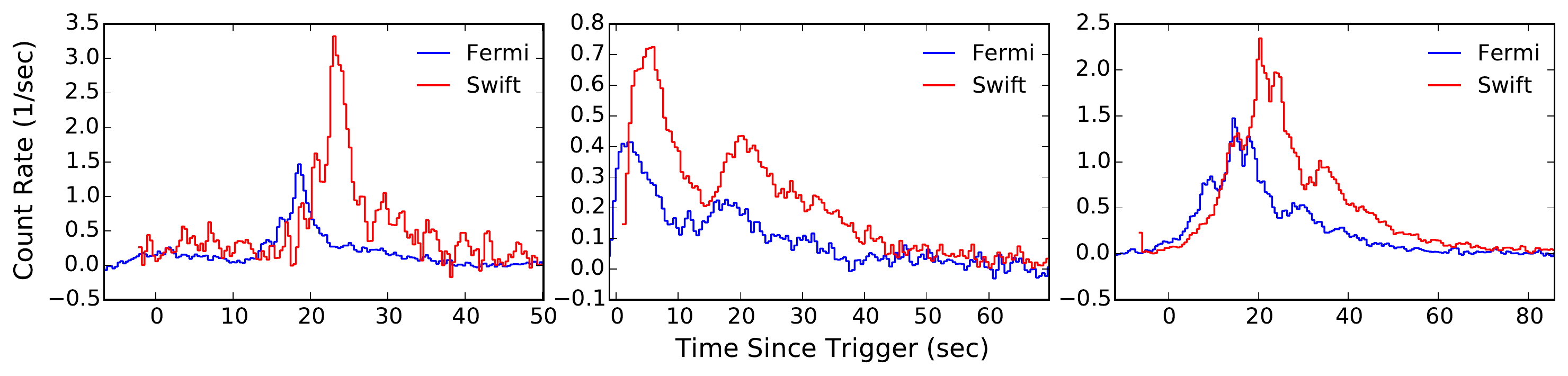}
  \label{fig:sf_lc}
\end{minipage}
\caption{\small 
    A gallery of Haar scaleograms $\sigma_{X, \Delta t}$, representing a variety of possible structure functions calculated for \textit{Fermi}/GBM and \textit{Swift}/BAT (both: 15--350 keV) with different level of sensitivity for detection of various GRBs.   The left, middle, and right panels correspond to GRB~110213A, GRB~080916A, and GRB~120119A, respectively. 
The first and second rows show the structure functions retrieved from the GRBs light curves detected by \textit{Fermi}/GBM and \textit{Swift}/BAT, respectively. The third row shows the light curves in the $T_{\rm 100}$ duration region.  
In each of these, the red dashed-lines represent a passage from the temporally-smooth ( $\sigma_{X, \Delta t} \propto \Delta t$) region to a flatter region and the red circle marks the extracted minimum variability timescale, $\Delta t_{\rm min}$, after which the light curves transition to a temporally-unsmooth behavior.  Triangles denote $3\sigma$ upper limits.
}
\label{fig:var}
\end{figure*}

\subsection{Consistency in the Joint \textit{Fermi}/GBM and \textit{Swift}/BAT Sample}
\label{sec:joint}

In Paper I, we studied the robustness of our minimum timescales extracted for simulated bursts as the S/N is varied.  It was demonstrated that the shapes of the $\sigma_{X,\Delta t}$ curves are highly stable as the S/N is strongly decreased (factor of ten), but the determination of the true $\Delta t_{\rm min}$ can be challenging.   This is because GRBs tend to show evidence for temporally-smooth variation between timescales of non-smooth variability (e.g., pulse rise times) -- which become harder to measure as S/N is decreased -- and the longer timescales associated with non-smooth variability (e.g., the full duration of the pulse).
The sample of bursts detected jointly  by both \textit{Swift}/BAT and \textit{Fermi}/GBM provides a rich dataset to study this behavior.  In addition to allowing us to verify consistency in the $\Delta t_{\rm min}$ estimates for bursts with similar S/N values, we can also directly observe (in many cases) the reliability of $\Delta t_{\rm min}$  for different S/N values.

Figure \ref{fig:var} captures the variety of scaleograms produced for bursts detected by both the \textit{Swift}/BAT and \textit{Fermi}/GBM instruments.
Here we utilize the 15--350 keV energy range for both \textit{Swift}/BAT and \textit{Fermi}/GBM, and we align the light curves and extract counts over the same time intervals for each burst.  Although the instruments do not have identical effective area curves in these ranges, choosing the same energy range should
minimize differences due to energy band (discussed in more detail in Section \ref{sec:hardness} below).

In the case of GRB~110213A (left panels), \textit{Fermi}/GBM captured the higher sensitivity burst light curve. Oppositely in the case of GRB~080916A (middle panels), \textit{Swift}/BAT captured a higher S/N light curve.
The S/N level can be gauged from the light curves and taken directly from the $\Delta t_{\rm S/N}$ values, with high S/N translating directly to lower $\Delta t_{\rm S/N}$.
There are many bursts (e.g., GRB~120119A, right panels) in the joint \textit{Fermi}/GBM and \textit{Swift}/BAT sample which correspond to closely similar S/N values and for which the resulting scaleograms are almost identical.  We note that minimum timescales based simply on $\Delta t_{\rm S/N}$ (e.g., \citealt{2000ApJ...537..264W}) directly track the noise floor level.  This is also the case for $t_{\beta}$, calculated according to the prescription of \citet{2013MNRAS.432..857M}.  In the most extreme examples (i.e., GRBs~090519A and 101011A), the $\Delta t_{\rm S/N}$ values differ by approximately an order of magnitude, the $t_{\beta}$ values
differ by approximately a factor of five, while the \tmin values are consistent (Table 2).
Our method distinguishes between the {\it minimum detectable} timescale and the true minimum timescale in a more robust (although not-perfect, as we discuss more below) fashion.

\begin{figure}[!ht]
\begin{center}
\includegraphics[width=.47\textwidth]{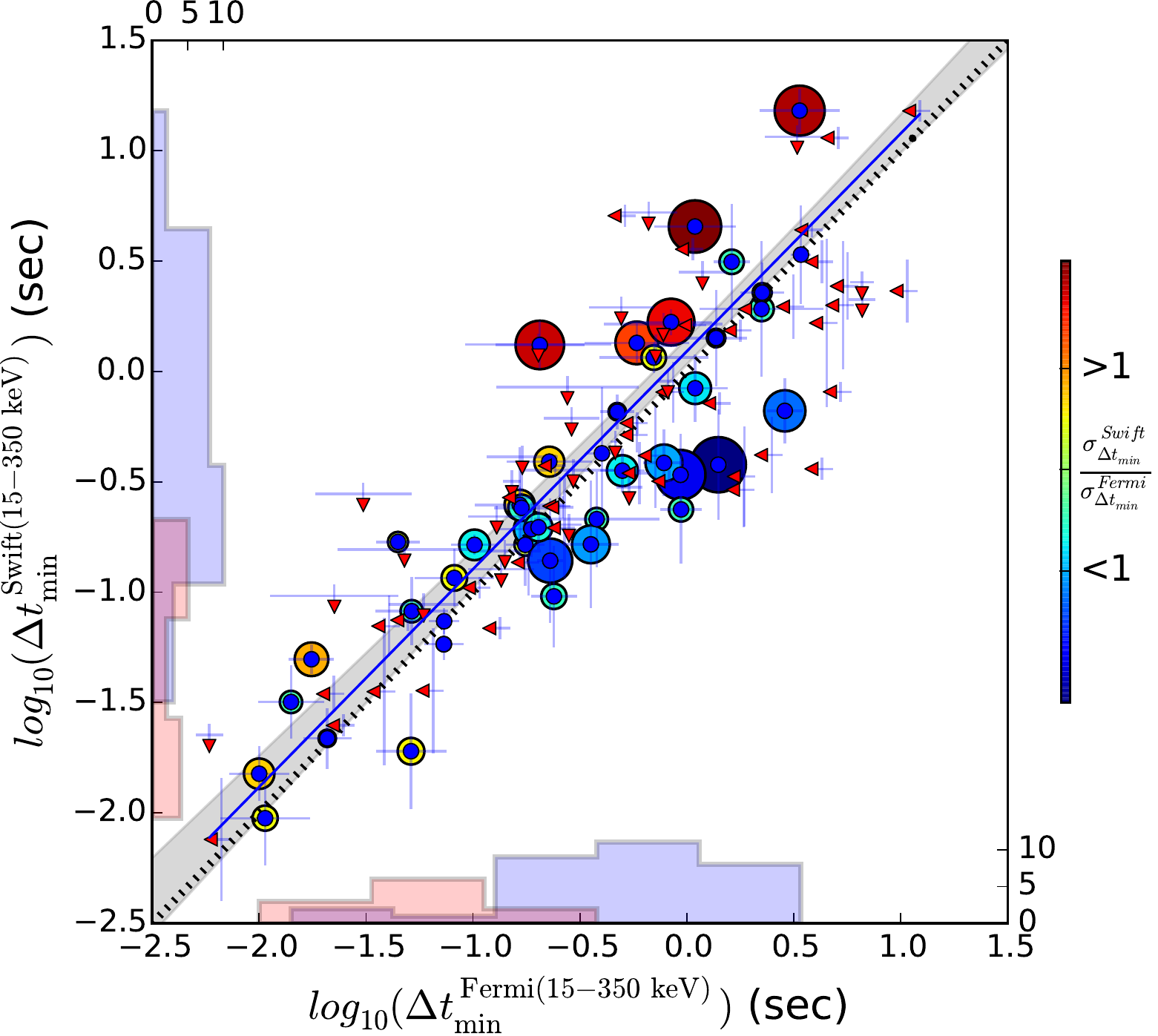}  
\caption{\small 
$\Delta t_{\rm min}$ for the sample of joint \textit{Fermi}/GBM and \textit{Swift}/BAT bursts. The red and blue histograms correspond to the short and long-duration GRBs, respectively. The arrows show the upper limit burst cases. The black dotted-line represents equality.
The relative size of the circles is scaled with the absolute value of the log of the ratio of flux variation at the shortest observable variability timescale $\Delta t_{\rm S/N}$, providing a measure of whether each satellite samples the same (small circles) or very different (large circles) regions of the scaleogram at the inferred $\Delta t_{\rm min}$.
The color bar can be used to identify which instrument generated the higher $\sigma_{X,\Delta t}$.
}
\label{fig:FS_scatter}
\end{center}
\end{figure}

Figure \ref{fig:FS_scatter} displays a scatter plot of $\Delta t_{\rm min}$ determined for \textit{Swift}/BAT versus \textit{Fermi}/GBM.
A line fit through the data points (blue curve with shaded gray 90\% confidence region) is consistent with the dotted-line representing equality.
The best-fit line has a normalization $= 1.13 \pm 0.13$ and a slope $= 0.99 \pm 0.02$.
For this fit the reduced $\chi^{2}  = 2.86$ (for 42 degrees of freedom) and is dominated by a small number of outliers.
The fraction of bursts not consistent with the fit, both below and above the line are: 12\% and 15\%, respectively.
The close consistency of this line with the unit line demonstrates that our method is robust and that our error bars, calculated by direct error propagation, are likely to be accurate.

We do note, however, that the \tmin values calculated for \textit{Swift} versus \textit{Fermi} do exhibit small, systematic differences.
On average, bursts detected by \swift (in the same energy band) tend to have 13\% longer \tmin values as compared to \textit{Fermi}.
Histograms showing the spread in the overall populations are also drawn along the axes in Figure \ref{fig:FS_scatter}.

To study the origin of the outliers to the fit in Figure \ref{fig:FS_scatter},  we scale the relative size of the circles with the absolute value of the log of the ratio of flux variation at the shortest observable variability timescale $\Delta t_{\rm S/N}$.  This is intended to provide an indication of whether each satellite sampled the same (small circles) or very different (large circles) regions of the scaleogram at the inferred $\Delta t_{\rm min}$.  The color bar can be used to identify which instrument generated the higher $\sigma_{X,\Delta t}$.

In general, we find that once the $\log({\Delta t_{\rm S/N}})$ ratios exceed 0.5 dex (corresponding to 0.5 dex in $\log({{\rm S/N}})$ or roughly a factor 10 in flux) the more sensitive satellite tends to yield a lower measurement of $\Delta t_{\rm min}$.  This is consistent with our findings from Paper I.  Given that such variation is not known a-priori in this case (because the light curves are not based on a simulation), the tendency to detect lower $\Delta t_{\rm min}$ when possible suggests a fractal nature of the phenomenon.  Care must be taken in interpreting GRB minimum timescales, because the phenomenology suggests these could always be limits on the true minimum timescales.  However, we do note the important feature of the scaleograms: hidden (i.e., low S/N) minimum timescales will always correspond to smaller variations in the fractional flux levels.   In this sense, a perfect accounting of the minimum timescales may not be necessary, because very short minimum timescales tend to represent fractionally tiny (or alternatively very rare) episodes in the GRB emission. \\

\begin{figure*}[!ht]
\begin{center}
\includegraphics[width=.91\textwidth]{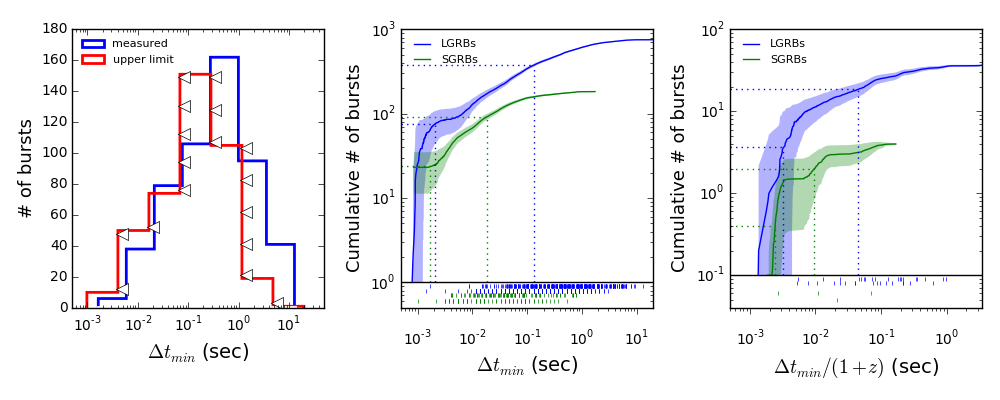}
\caption{\small 
Left panel: the histograms of $\Delta t_{\rm min}$ with measurements (blue) and for GRBs allowing for upper limits only (red). Middle and right panels: the cumulative histograms of bursts in the observer and source frames, respectively.  The KM estimation curve with $1\sigma$ error region around the curve is shown in these panels. The dotted lines correspond to the minimum timescale of the lowest 10\% and 50\% of bursts, shown for the short and long-duration GRBs, separately. 
Sub-panels show the locations of detections and upper-limits, as in Figure \ref{fig:km}.  For long-duration (short-duration) GRBs, we have 421 (107) measurements and 334 (76) upper limits in the observer frame and 24 (3) measurements and 18 (1) upper limits in the source frame.
}
\label{fig:hist}
\end{center}
\end{figure*} 

\subsection{Distribution of $\Delta t_{\rm min}$ Values for \textit{Fermi}/GBM}

Figure \ref{fig:hist} (left) shows histograms for the \fermi GRBs permitting measurement of and also upper limits on $\Delta t_{\rm min}$.
The two distributions have consistent mean values.
The middle and right panels of Figure \ref{fig:hist} show the KM cumulative histograms in the observer and source frames, respectively.
The dotted-lines correspond to the minimum timescale of the lowest 10\% and 50\% (median) of short and long-duration bursts.

We find a median minimum timescale for long-duration (short-duration) GRBs in the observer frame of 134 ms (18 ms).  In the source frame, we find a median minimum timescale for long-duration (short-duration) GRBs of 45 ms (10 ms).  It is interesting that these numbers are a factor of 3--10 smaller than those we found for \swift in Paper I.  The largest differences, in the case of short-duration GRBs,  are attributable to the increased number of well-detected short-duration GRBs by \textit{Fermi}.  As we discuss below (Section \ref{sec:hardness}), $\Delta t_{\rm min}$ also appears to vary by a factor of $\approx 3$ depending on the burst hardness.  The \textit{Fermi} sample is studied using the full energy range, and the sample appears to be spectrally harder than the \textit{Swift} sample, overall.

We also report \tmin of the most exotic GRBs in \fermi sample -- the lowest $10^{\rm th}$ percentile of bursts with the shortest $\Delta t_{\rm min}$. The $10^{\rm th}$ percentile \tmin values for long-duration (short-duration) GRBs in the observer frame found to be 2.2 ms (1.9 ms).   In the source frame, we find 2.9 ms (2.4 ms).  These numbers are consistent with the findings in Paper I that millisecond variability appears to be rare in GRBs.

From Figure \ref{fig:hist}, we find that the $\Delta t_{\rm min}$ distribution of long-duration GRBs is displaced from that of short-duration GRBs ($16 \sigma$, $t$-test $17 \sigma$, log-rank test \citep{pmid5910392}).
The log-rank test includes the upper limits, unlike the $t$-test.
This finding is consistent with the presented results in Paper I for \textit{Swift}.
This discrepancy is still present in the source frame ($2.3 \sigma$, $t$-test and $3.4 \sigma$, log-rank test) unlike in Paper I where the distribution centers appeared to be consistent. The \swift small sample of short-duration GRBs with known-$z$ is likely the main reason for the observed degeneracy. The significant observer frame discrepancy is likely driven by the fact that short-duration GRBs tend to be detected only at low-redshift, unlike long-duration GRBs which span a broad range of redshifts.
Examining the dispersion in $\log(\Delta t_{\rm min})$ values, we see no strong evidence for dissimilar values for the long and short-duration samples ($<1.3\sigma$, $F$-test).   This finding is also fully consistent with the presented results in Paper I, where it was also found (using a sample of \swift GRBs) that the two histograms are quite broad and very similar in dispersion.

\begin{figure}[ht]
\begin{center}
\includegraphics[width=.47\textwidth]{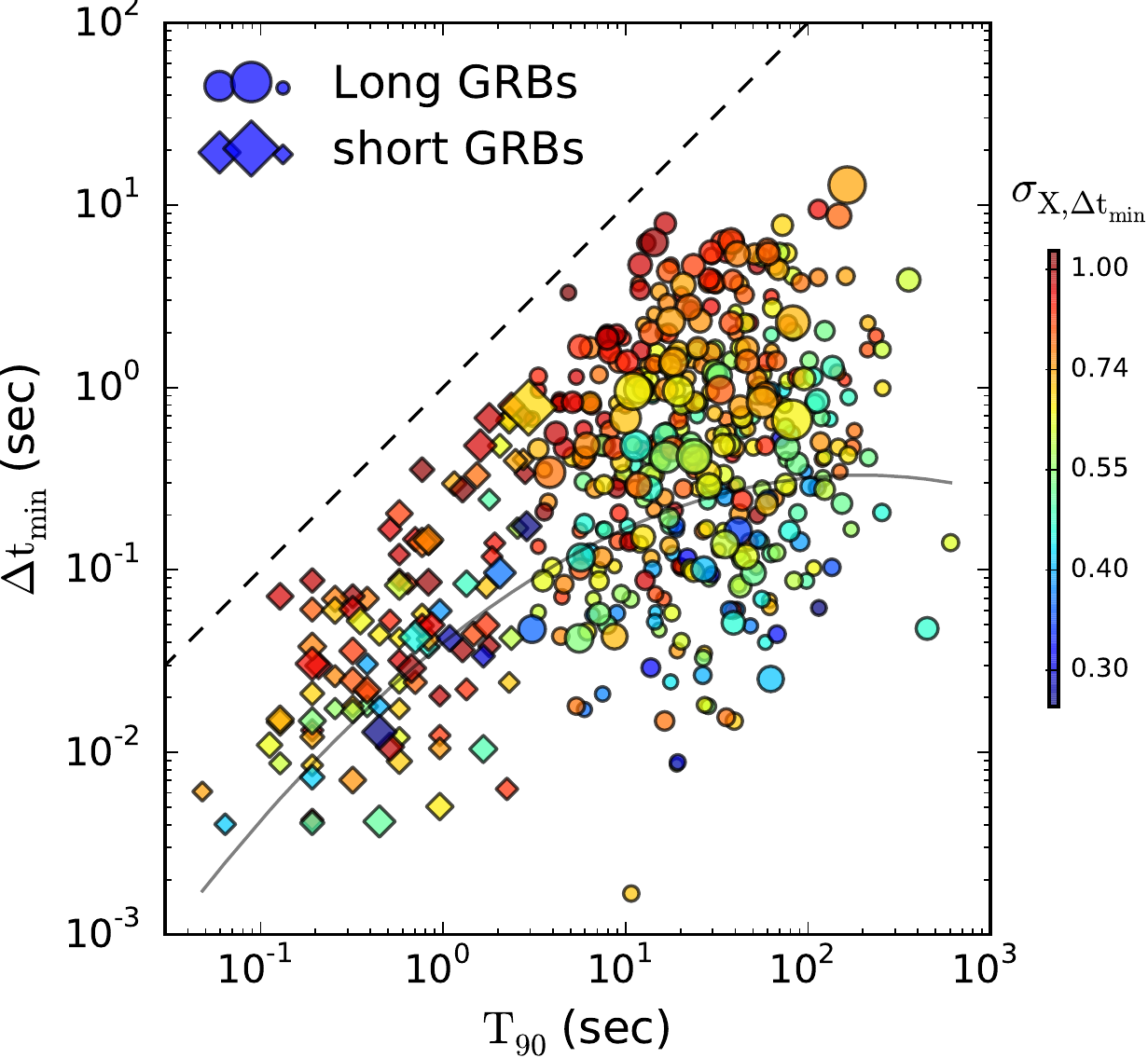}  
\caption{\small 
The GRB minimum timescale, $\Delta t_{\rm min}$, plotted versus the GRB $T_{\rm 90}$ duration. Circles (diamonds) represent long-duration (short-duration) GRBs. The point colors represent the flux variation level ($\sigma_{X, \Delta t_{\rm min}}$) at $\Delta t_{\rm min}$. Also plotted as a curved line is the typical minimum observable timescale, $\Delta t_{\rm S/N}$, as a function of $T_{\rm 90}$. The symbol sizes are proportional to the ratio of $\Delta t_{\rm min} / \Delta t_{\rm S/N}$ for each GRB. The dashed line shows the equality line. 
} 
\label{fig:tmin_vs_T90}
\end{center}
\end{figure}

Figure \ref{fig:tmin_vs_T90} displays our minimum variability timescale, $\Delta t_{\rm min}$, versus the GRB duration, \tninty. The short and long-duration GRBs are shown with diamond and circle symbols, respectively.  In this plot the relative size of symbols is proportional to the ratio between minimum variability and S/N timescale ($\Delta t_{\rm min} / \Delta t_{\rm S/N}$).  As described above, $\Delta t_{\rm S/N}$ represents the first statistically significant timescale in the Haar wavelet scaleogram.  The color of the points in Figure \ref{fig:tmin_vs_T90} corresponds to the flux variation level, $\sigma_{X,\Delta t}$, at $\Delta t_{\rm min}$.   A curved black line is also plotted to show a typical value for the minimum observable time ($\Delta t_{\rm S/N}$) versus $T_{\rm 90}$.
Values for $T_{\rm 90}$ are taken from Table 7 of  \citet{2014ApJS..211...13V}.

We first note from the colors in Figure \ref{fig:tmin_vs_T90} that GRBs with $\Delta t_{\rm min}$ close to \tninty\, tend to have flux variations of order unity.  These are bursts with simple, single-pulse time profiles.  As can be seen from the range of point sizes in Figure \ref{fig:tmin_vs_T90}, most are not simply low S/N events where fine time structure cannot be observed.  Also, we see that there are GRBs with both high and low S/N which have complex time-series ($\Delta t_{\rm min} \ll T_{\rm 90}$).  Based on the point sizes, the short-timescale variation have higher ratio of $\Delta t_{\rm min} / \Delta t_{\rm S/N}$ for the short-duration GRBs of the similar \tmin in comparison with that of the long-duration GRBs. 
Short-duration GRBs tend to have a higher $\sigma_{X,{\Delta t}}$ for the similar value of \tmin compared with the long-duration GRBs. \\
These findings are all consistent with the similar results explained in Paper I; although we have a better ratio of short-duration GRBs to long-duration GRBs, here. 

From a Kendall's $\tau$-test \citep{kendall1938new}, we find only marginal evidence that $\Delta t_{\rm min}$ and $T_{\rm 90}$ are correlated ($\tau_k = 0.33$, $11\sigma$ above zero). 
The $\Delta t_{\rm min}$ values in Figure \ref{fig:tmin_vs_T90} are bound from above by $T_{\rm 90}$, and they do not strongly correlate with $T_{\rm 90}$ within the allowed region of the plot.  In Paper I, we studied this relation for the entire sample of \swift GRBs and found only a marginal evidence that \tmin and $T_{\rm 90}$ are correlated ($\tau_k = 0.38$, $1.5\sigma$). Even when we utilized the robust duration  estimate $T_{\rm R45}$ \citep{2001ApJ...552...57R} in place of $T_{\rm 90}$ no significant correlation was found ($\tau_k = 0.6$, $2.4\sigma$). 
If we perform a truncated Kendall's $\tau$ test which only compares GRBs above one-another's threshold \citep{2002ApJ...565..182L}, the correlation strength drops precipitously ($\tau_k = 0.06$, $1.4\sigma$). We, therefore, believe there is no strong evidence supporting a real correlation between $\Delta t_{\rm min}$ and $T_{\rm 90}$.

\begin{figure}[!ht]
\begin{center}
\includegraphics[width=.47\textwidth]{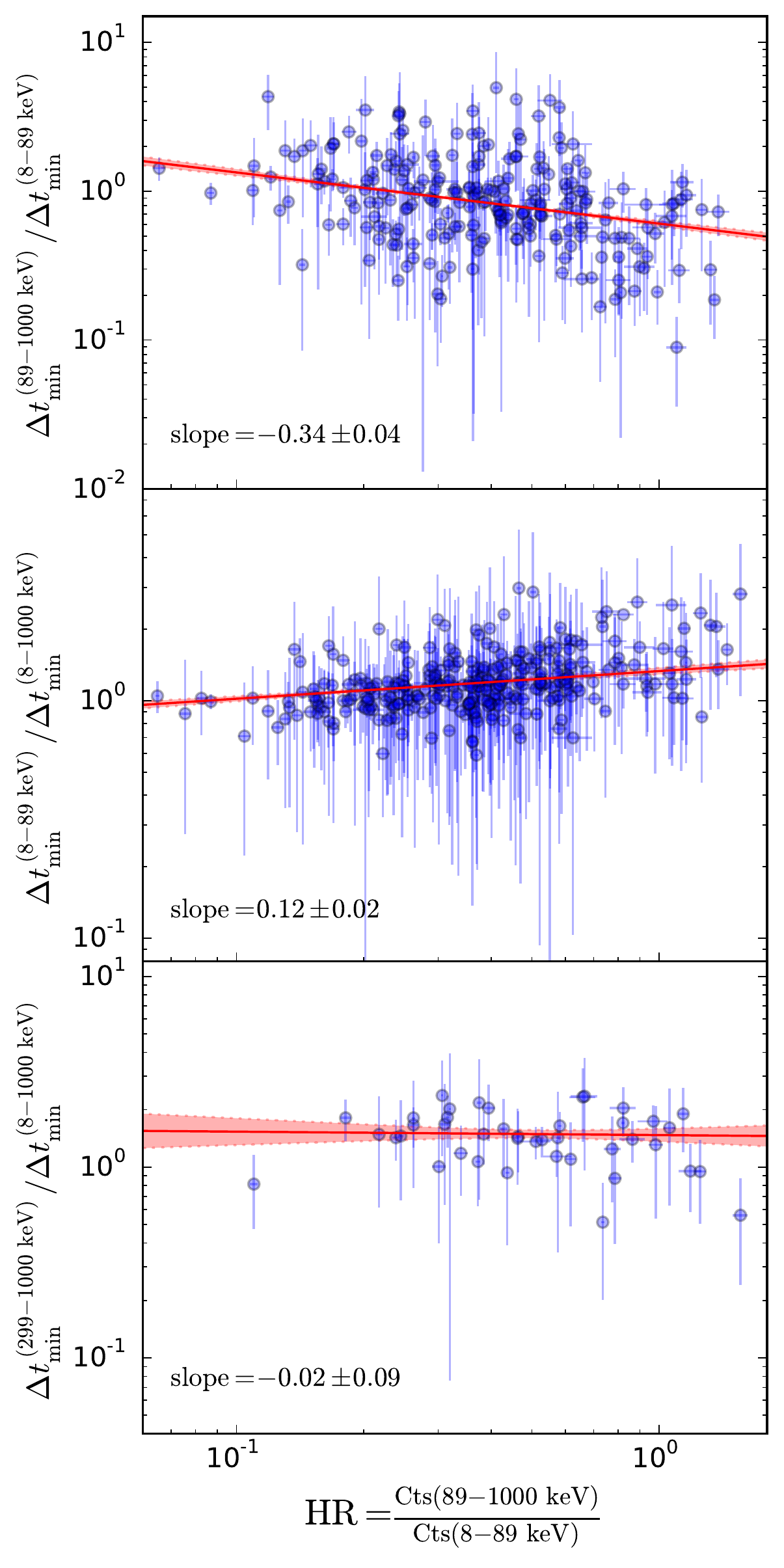} 
\caption{\small
Top panel: The ratio of minimum variability timescale for channels $3+4$ and channels $1+2$, plotted against hardness ratio for the corresponding composite channels.  
Middle and bottom panels: The ratio of \tmin for channels $1+2$ and channel 4 over full energy band, separately plotted against hardness ratio.
 The best fitted linear model through the bursts including shaded $1\sigma$ error region is also shown in each panel. 
}
\label{fig:hardness}
\end{center}
\end{figure}

\subsection{The Dependence of \tmin on Spectral Hardness}
\label{sec:hardness}

We investigate here how a burst's spectral hardness impacts its minimum variability timescale.  We define the hardness ratio (\textit{HR}) as the total counts in the hard composite channel (89--1000 keV, our combined channels 3 and 4) divided by the total counts in the soft composite channel (8--89 keV, our channels 1 and 2).
We plot in Figure \ref{fig:hardness} (top panel) the ratio of \tmin for these two composite channels 
against the \textit{HR} of the two corresponding bandpasses.
GRBs with harder spectra tend to have a lower \tmin ratio, by as much as a factor $\approx 3$, for both short and long-duration GRBs.
This relationship can be captured using a best-fitted linear model through all the bursts, shown in Figure \ref{fig:hardness} (top panel), with slope $= -0.34 \pm 0.04$.

The change in minimum timescale with hardness can be understood from the effects of relativistic beaming on emission instantaneously emitted in the rest frame by a moving shell \citep[e.g.,][]{1996ApJ...473..998F,2002ApJ...578..290R,2003ApJ...596..389K}.  If the material on the line-of-site has a Doppler factor $\Gamma (1-\beta)$
, propagating with a speed $v = \beta c$ and Lorentz factor $\Gamma$,
material above or below the line of site at angle $\theta$ will have a Doppler factor $\Gamma (1-\beta cos(\theta)) \approx (1+(\Gamma \theta)^2)/2 \Gamma$, larger by a factor $1+(\Gamma \theta)^2$.  The off-axis emission will also arrive later, at a time $t-t_e = R/c (1-\cos(\theta))$, where $R$ is the emission radius, after the start of the emission at $t_e$.  If we assume $R = 2 \Gamma^2 c t_e$, then the Doppler factor increases in time, in the observer frame, as $t/t_e$.  As a result, the photon flux observed at fixed energy $E$ will decrease as higher and higher rest-frame-energy photons reach the bandpass, as $(t/t_e)^{\alpha-2}$.  Here, $\alpha$ is the photon index and the power of 2 arises from relativistic beaming.

Thus, we expect that impulsive releases of energy in the rest frame will be smoothed over -- in a fashion that is stronger at low energy ($\alpha \approx -1$) as compared to high energy (above $E_{\rm peak}$, $\alpha \lessim -2$) -- as viewed in the observer frame.  The degree of smoothing expected above $E_{\rm peak}$ is a factor 2--3 less than the smoothing expected at observer frame energies below $E_{\rm peak}$.  This effect naturally explains the decreasing minimum timescale we observe with increasing spectral bandpass, and it suggests that the tightest constraints on minimum timescale should be obtained from the highest available instrument bandpass.  It should also be sufficient to confirm that $E_{\rm pk}$ is below, or perhaps within, a given bandpass. 

Figure \ref{fig:hardness} (middle panel) shows the ratio of \tmin for the soft composite channel over the full energy band against the \textit{HR}. This plot 
shows how \tmin is approximately the same in each bandpass until the hardness ratio goes beyond roughly its median value. The bursts in this plot well-fitted
by a line with slope $= 0.12 \pm 0.02$.

The ratio of \tmin for the hardest channel (\#4) over the full energy band against the \textit{HR} is shown in Figure \ref{fig:hardness} (bottom panel).
Here, the best-fit line (slope $=-0.02 \pm 0.09$) is consistent with being flat: the minimum timescales appear to be independent of this hardness ratio for all but perhaps the hardest handful of \fermi GRBs.
We conclude that utilizing the full \textit{Fermi}/GBM bandpass -- which yields \tmin constraints consistent with those derived from the soft energy channel for soft GRBs and also \tmin constraints consistent with those derived from the hard energy channel for hard GRBs -- is an acceptable procedure for determining the tightest constraints on $\Delta t_{\rm min}$.

\subsection{Constraints on the Size of the Central Engine}
\label{sec:prog}

The minimum timescale provides an upper limit on the size of the GRB emission region, in turn providing hints on the nature of the GRB progenitor and potentially shedding light on the nature of emission mechanism.
In Paper I, we summarized how an association of a minimum timescale with a physical size is not unique, because the observed timescales depend strongly also on the emitting surface velocity.

The minimum Lorentz factor $\Gamma$ can be estimated from the compactness argument \citep{2001ApJ...555..540L}.  
If we assume a spectrum with photon index $\alpha=-2$ (see, \citealt{2013ApJS..209...11A}, Figure 25) -- typical for GRB spectra above the pair-production limit and also appropriate for the range of energies which dominate the luminosity (near the $\nu F_{\nu}$ spectral peak) -- we find
\be
\Gamma \gtrsim 110 \, \left (\frac{L}{10^{51} \, \rm erg/s} \, \frac{1+z}{\Delta t_{\rm min} / 0.1 \, \rm sec } \right )^{1/5},
\ee
where $L$ is the gamma-ray luminosity.  
If we regard $\Delta t_{\rm min}$ as corresponding to the bolometric emission, it is most natural to use the full \textit{Fermi}/GBM bandpass for its estimation rather than a fixed rest frame bandpass.  It could be argued that corrections should also be made to account for spectral hardness, based perhaps on the assumption that GRBs have a single, fixed rest frame hardness -- an unlikely possibility -- modulated only by Lorentz factor.  However, based on the analysis in Section \ref{sec:hardness} above, any corrections would be small.

Utilizing our \tmin estimates and limits for the full \textit{Fermi}/GBM bandpass, we find that 50\% of \fermi GRBs must have $\Gamma>190$.  In the case of the most energetic events, 10\% of \fermi GRBs require $\Gamma > 410$.  To calculate these fractions for short-duration bursts without measured redshift, we follow \citet{2014MNRAS.442.2342D} in assigning an average $z=0.85$.
For long-duration GRBs lacking redshift, we assign the average $z=2.18$.

Similarly, for some maximally allowed $\Gamma_{\rm max}$, compactness limits the emission radius to be greater than
\be
R_{\rm min} \simeq 2.8 {\times} 10^{10} \, \frac{L}{10^{51} \, \rm erg/s} \left (\frac{\Gamma_{\rm max}}{1200} \right )^{-3} \, \rm cm. \\
\ee
This minimum bound on the radius can be compared to the maximum bound on the radius established by the temporal variability:
\begin{align}
R_{\rm max}  &= c \, \frac{\Delta t_{\rm min}}{1+z} \, \Gamma_{\rm max}^2, \nonumber \\
&\simeq 4.4 {\times} 10^{15} \, \frac{\Delta t_{\rm min} / 0.1 \, \rm sec }{1+z} \, \left (\frac{\Gamma_{\rm max}}{1200} \right )^2 \, \rm cm.
\end{align}
Here, we conservatively take $\Gamma_{\rm max} \sim 1200$ from \citet{2011ApJ...738..138R}.  

If emission were to occur at the minimum allowable radius, $R_{\rm min}$, it would
correspond to variability timescales as short as $\Delta t = R_{\rm min} / (2c \Gamma_{\rm max}^2) \lessim 1 \mu$s.  Because such timescales are
not observed, a more realistic bound on the minimum emission radius is $R_c = 2c \Gamma_{\rm min}^2 \Delta t_{\rm min}/(1+z)$, or
\be
R_c \simeq 7.3 {\times} 10^{13} \, \left (\frac{L}{10^{51} \, \rm erg/s} \right )^{2/5} \left (\frac{\Delta t_{\rm min} / 0.1 \, \rm sec }{1+z} \right )^{3/5} \, \rm cm.
\label{eq:ri}
\ee

\begin{figure}[ht!]
\begin{center}
\includegraphics[width=.47\textwidth]{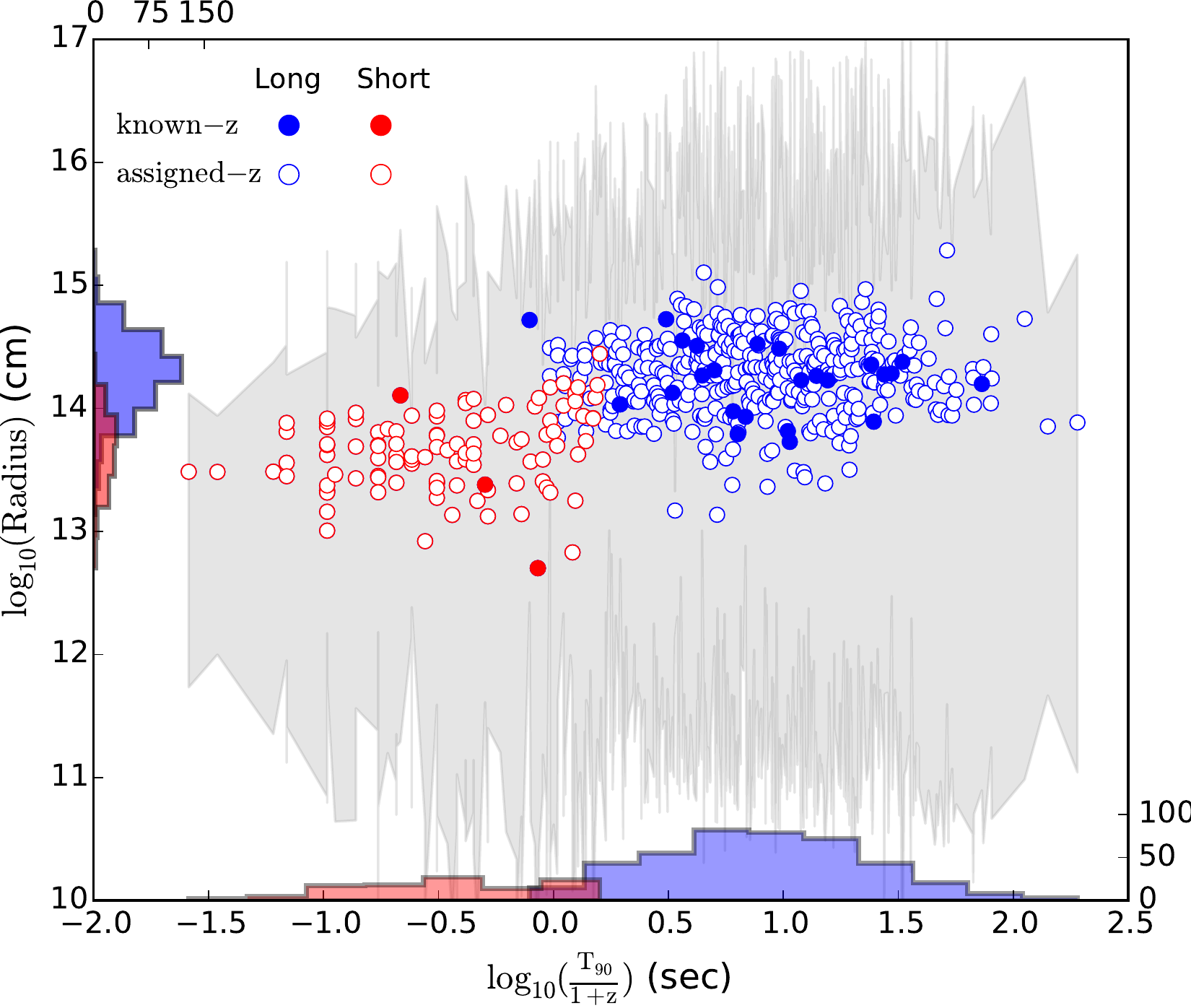}
\caption{\small
The characteristic emission radii $R_c$ (Equation \ref{eq:ri}) plotted versus rest frame $T_{\rm 90}$ for the \textit{Fermi}/GBM bursts. 
The shaded region shows the interval between the minimum and maximum emission radii allowed.
The bursts with known-$z$ and assigned-$z$ are denoted with filled and unfilled circles, respectively.  
The short and long-duration GRBs are denoted with red and blue colors, respectively.
}
\label{fig:radius}
\end{center}
\end{figure}

Figure \ref{fig:radius} shows the emission radius, $R_c$, for all the bursts with measured \tmin in \textit{Fermi}/GBM sample versus rest-frame $T_{\rm 90}$.  The shaded region shows the interval between the $R_{\rm min}$ and $R_{\rm max}$.  The interpretation of $R_c$ as a characteristic minimum radius for the emission is motivated further in Section \ref{sec:Conclude}.

The short-duration GRBs have a KM mean $R_c=3.3 \times 10^{13}$ cm.  This is about four times smaller than the KM mean $R_c = 1.3 \times 10^{14}$ cm for long-duration GRBs.
While this represents a statistically significant separation ($18 \sigma$, $t$-test), it is substantially less than the factor of approximately twenty separation between the mean $T_{90}$ durations \citep[Figure \ref{fig:radius}, also][]{1993ApJ...413L.101K}.  In contrast to the
findings of \citet{2014ApJ...794L...8B} -- where the emission radius was argued to simply scale with the $T_{\rm 90}$ duration -- we find a broader overlap in the populations.

\subsection{Evolution of $\Delta t_{\rm min}$ with $z$}
\label{sec:cosmic_time}

Because GRBs are present over a very broad redshift range, the signature of time-dilation -- and perhaps of any evolution in GRB time-structure with redshift -- should be present in GRB time-series.   Finding the signature of time-dilation in GRBs has remained elusive (\citealt{1994ApJ...424..540N}; \citealt{2013ApJ...765..116K}, but see, e.g., \citealt{2013ApJ...778L..11Z}). In our previous attempt described in Paper I, we utilized \swift GRBs and demonstrated a correlation between \tmin and redshift, marginally stronger than expected simply from time-dilation.  We discussed how this excess correlation strength was possibly due to the utilization of a fixed observer frame bandpass instead of a fixed rest frame bandpass in the analysis.  For \textit{Fermi}/GBM, the broad instrument energy range permits analysis in a fixed rest frame bandpass.

\begin{figure}[ht!]
\begin{center}
\includegraphics[width=.50\textwidth]{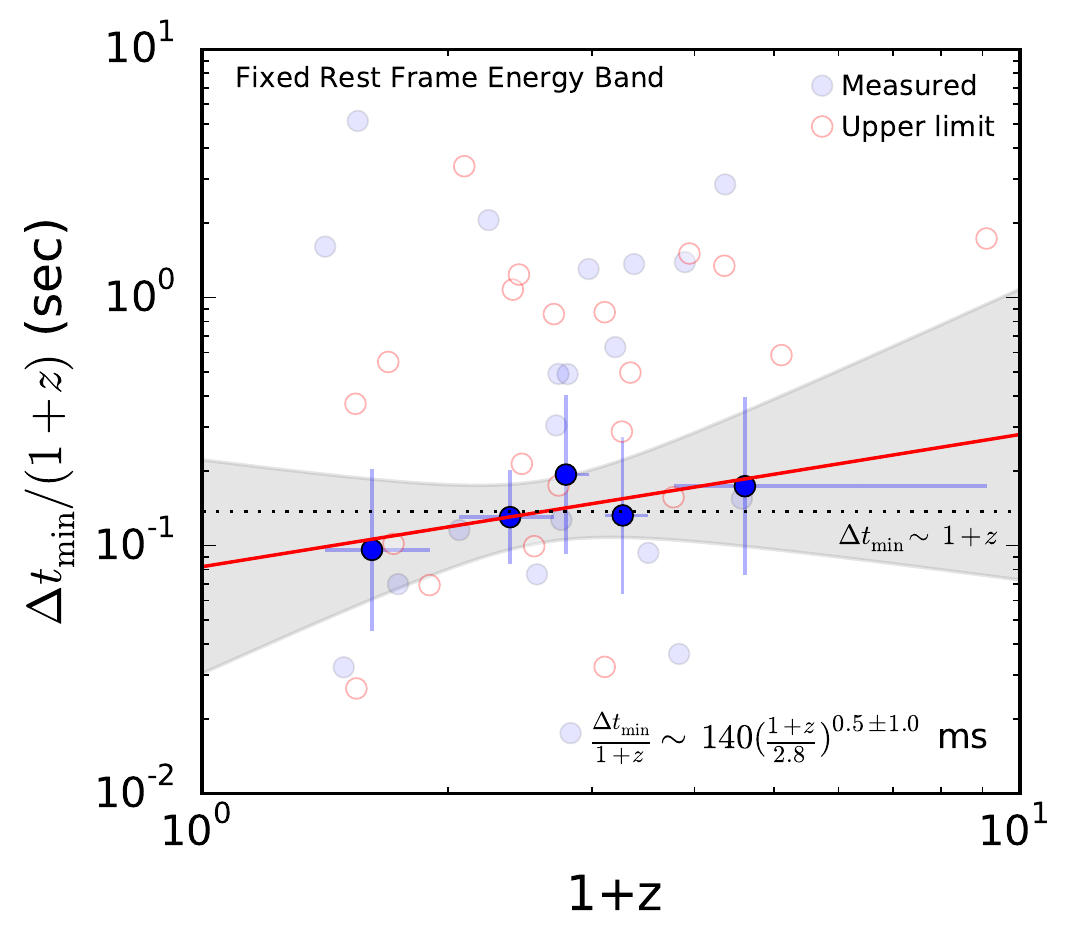}
\caption{\small
Minimum variability timescale in the rest frame 89--299 keV energy band versus redshift, $z$.  The blue circles show the KM mean values of \tmin for groups of 7--10 bursts of similar redshift. The shaded region represents the $1\sigma$ confidence interval for the fitted red line.
The dotted black line shows the expected evolution due to simple cosmological time dilation, namely $\Delta t_{\rm min} \sim 1+z$.
The faint blue circles show all GRBs with measured $\Delta t_{\rm min}$ and known-$z$ and the unfilled circles show GRBs with upper limit values for \tmin.
}
\label{fig:time_dilation_a}
\end{center}
\end{figure}

We identify 46 \textit{Fermi} GRBs, including 4 short-duration GRBs, with measured redshifts.  Light curves are extracted in the rest frame 89--299 keV band and analyzed.
In Figure \ref{fig:time_dilation_a} we plot $\Delta t_{\rm min} / (1+z)$ versus $1+z$ for the long-duration GRBs.  Redshift values are taken from \citet[and references therein]{2007ApJ...671..656B,2010ApJ...711..495B}, \citet{2013AstRv...8a.103B}, and this webpage\footnote{\url{http://www.mpe.mpg.de/~jcg/grbgen.html}}.
The blue circles in Figure \ref{fig:time_dilation_a} correspond to the KM mean values of \tmin for sets of between 7 and 10 bursts, grouped by redshift intervals.
The unbinned data are plotted in the background for the entire sample and for those with measured \tmin using unfilled and filled circles, respectively.
We find that the binned data can be well-fitted by a line $\Delta t_{\rm min} / (1+z) \sim 140 ((1+z ) /2.8)^{0.5\pm1.0}$ ms, suggesting possibly increase in timescale with $z$ but also consistent the prediction of simple time-dilation (dotted line).

\section{Conclusions}
\label{sec:Conclude}
Using a technique based on Haar wavelets, previously developed in Paper I, we studied the temporal properties of a large sample of GRB gamma-ray prompt-emission light curves captured by the GBM instrument onboard \fermi prior to July 11, 2012.  We analyzed the time histories in four energy bands.  While the derived values for $\Delta t_{\rm min}$ are highly-dependent upon bandpass, we find that the use of the full energy band allows for the tightest constraints on the size of the emission region.  In principle, the highest-energy bandpass should yield the tightest constraint (Section \ref{sec:hardness}).  However, S/N in the highest-energy channels is often low; the full energy bandpass allows for increased S/N while maintaining a consistent $\Delta t_{\rm min}$ estimate.

Applying our technique to the joint \textit{Fermi}/GBM and \textit{Swift}/BAT sample, we find close consistency in the minimum timescales derived for each instrument.
However, as suggested by simulations in Paper I -- and observed for a handfull of bursts of widely varying S/N in Section \ref{sec:joint} -- $\Delta t_{\rm min}$ values below the measurement limit ($\Delta t_{\rm S/N}$) can be present.  It is thus important to consider our $\Delta t_{\rm min}$ values as defined given the observed data, with the possibility of improved limits given better data.
We urge caution, in particular, in interpreting minimum timescales determined using hard X-ray data (e.g., \textit{Swift}/BAT).
Minimum timescale estimates using the full \textit{Fermi}/GBM bandpass are a factor 2--3 times more constraining than those determined from \textit{Fermi}/GBM data in a \textit{Swift}/BAT bandpass.

Considering measurements and limits, we find a median minimum variability timescale in the observer frame
of 134 ms (long-duration; 18 ms for short-duration GRBs).   In the source frame, for a smaller sample of 33 GRBs, we find a median timescale of 45 ms (long-duration; 10 ms for short-duration GRBs).
This finding validates our previous results in Paper I, confirming that millisecond variability appears to be rare in GRBs.
In the most extreme examples, 10\% of the long-duration GRB sample yields evidence for 2.2 ms variability (1.9 ms for short-duration GRBs).  
In the source frame, we find similar numbers, 2.9 ms for long-duration GRBs and 2.4 ms for short-duration GRBs.
Even if we restrict to the 67 GRBs within minimum detectable timescales $t_{\rm S/N}<10$ ms, only 10\% of the brightest and/or most impulsive GRBs show evidence for variability on timescales below 4.2 ms in the observer frame.

\subsection{Constraints on the Fireball Model}

In the ``external shock'' model \citep[e.g.,][]{1992MNRAS.258P..41R}, gamma-rays are produced as the GRB sweeps up and excites clouds in the external medium.
The extracted $\Delta t_{\rm min}$ can circumscribe the size scale of the impacted cloud along the line of sight. For a thin shell \citep[e.g.,][]{2006RPPh...69.2259M}, the gamma-ray radiation will start when the relativistic shell hits the inner boundary of the cloud with the peak flux produced as the shell reaches the densest region or center of the cloud. The size scale of the impacted cloud is limited  by $2 \Gamma^2 c\,\Delta t_{\rm min}$ since the shock is moving near light speed \citep{1996ApJ...473..998F}. For the smallest $\Delta t_{\rm min}$ found ${\sim} 1$ ms,  and assuming $\Gamma < 1000$, the cloud size must be smaller than $4$ AU. \par

If the angular size of an impacted cloud as viewed from the GRB central engine is $\Theta$,
the minimum variability timescales is constrained to be $\delta \Theta \, \Gamma < \Delta t_{\rm min}/ 2 \, T_{\rm rise}$ (Paper I).
Here, $T_{\rm rise}$ denotes the overall time to reach the maximum gamma-ray flux.
The fraction of the emitting shell that becomes active is expected to be of order $0.1 \Delta t_{\rm min}/ 2 \, T_{\rm rise}$ \citep{1999ApJ...512..683F}.
For the bursts in the \fermi sample with typical minimum variability timescale $\Delta t_{\rm min} \sim T_{\rm rise}$,
there is no need to consider a highly-clumped external medium and the external shock scenario is viable.

However, there are many bursts (e.g., Figure \ref{fig:tmin_vs_T90}) which do exhibit $\Delta t_{\rm min}/ T_{\rm rise} \ll 1$.
If this variability results from a clumped external medium, then a significant fraction of the energy from the GRB must escape without interacting
and producing gamma-rays.  Early X-ray afterglow observations \citep[e.g.,][]{2006ApJ...642..389N}, on the other hand, demonstrate the need for a high (order unity) efficiency in tapping the
kinetic energy of the flow to produce gamma-rays.  Thus, external shocks likely cannot explain the finest-time-scale variability.

In the ``internal shock'' scenario \citep[e.g.,][]{1994ApJ...430L..93R}, the relativistic expanding outflow released from a central engine is assumed to be variable, consisting of multiple shells of different $\Gamma$.  The dispersion in $\Gamma$ is related to the observed variability of the light curve, as $\Delta \Gamma / \Gamma \approx 1/2 \, (\Delta t_{\rm min} / T_{\rm rise})$ (Paper I), with
many of the \textit{Fermi} light curves requiring $\Delta \Gamma \approx \Gamma$.
Efficient production of gamma-rays also requires $\Delta \Gamma \approx \Gamma$ \citep{1999PhR...314..575P,2001ApJ...551..934K}.
It is, therefore, natural to assume that some of the gamma-ray emission is released with the minimum possible Lorentz factor $\Gamma_{\rm min} \approx 200$ (Section \ref{sec:prog}) allowed from compactness considerations.
As a result, considering variability at the few millisecond level, some GRBs must emit at radii of order $R_c \approx 2\Gamma_{\rm min}^2 c\,\Delta t_{\rm min} \approx 10^{13}$ cm (Equation \ref{eq:ri}, Figure \ref{fig:radius}).  This is also the extent to which minimum variability timescales can limit the size of the progenitor.

We find that long-duration GRBs appear to have typical emission radii $R_c\approx 1.3{\times} 10^{14}$ cm, while short-duration GRBs have four times smaller typical emission radii, $R_c\approx 3.3{\times} 10^{13}$ cm.  There is large scatter in the inferred radii of each population, and the distributions appear to strongly overlap.  It is unclear whether the dichotomy in short and long-duration GRB $T_{\rm 90}$ durations maps cleanly to a similar dichotomy in the size of the emission regions.

Finally, we note that our minimum timescales appear to correlate with redshift in fashion consistent with cosmological time-dilation.  Correcting for this, we find no significant evidence that $\Delta t_{\rm min}/(1+z)$ evolves with redshift.  This may be partly because the number of \fermi GRBs with measured redshifts is low (e.g., as compared to \textit{Swift}; Paper I).  Future increases in the sample size will surely allow for tighter constraints on minimum emission radii, Lorentz factors, and progenitor dimensions as well as allowing us to better understand whether any of these quantities vary with cosmic time.


\clearpage
\clearpage


 \vfill
\eject
\clearpage
\begin{center}
\begin{scriptsize}
\begin{longtable}{@{}l@{}c@{}c@{}c@{}c@{}c@{}c@{}c@{}c@{}c}

\caption[]{GRB Minimum Timescales} \label{tab:spec_table} \\[0.5ex]

\hline \hline \\[2ex]
   Trigger ID & \; GRB Name & $\Delta t_{\rm min}$ & $\Delta t_{\rm S/N}$ & $T_{\rm 90}$ & $T_{\rm 100}^{\rm \,start}$ & $T_{\rm 100}^{\rm \,stop}$ & $\sigma_{X, t_{\rm min}}$ & $\sigma_{X, t_{\rm S/N}}$ & z \\[0.5ex]
    &  & (s) & (s) & (s) & (s) & (s) &  &  &  \\[0.5ex] \hline
   \\[1.8ex]
\endfirsthead

\multicolumn{10}{l}{{\tablename} \thetable{} -- Continued} \\[0.5ex]
  \hline \hline \\[2ex]
   Trigger ID & \; GRB Name & $\Delta t_{\rm min}$ & $\Delta t_{\rm S/N}$ & $T_{\rm 90}$ & $T_{\rm 100}^{\rm \,start}$ & $T_{\rm 100}^{\rm \,stop}$ & $\sigma_{X, t_{\rm min}}$ & $\sigma_{X, t_{\rm S/N}}$ & z \\[0.5ex]
    &  & (s) & (s) & (s) & (s) & (s) &  &  &  \\[0.5ex] \hline
  \\[1.8ex]
\endhead

  \\[1.8ex] \hline
  \multicolumn{10}{l}{{Continued on Next Page\ldots}} \\[0.5ex]
\endfoot

  \\[1.8ex] \hline \hline
\endlastfoot
080714086 &  080714B &  $0.821\pm0.223$ &  $0.522\pm0.045$ &  $5.376\pm2.360$ &  -3.437  & 7.283 &   0.90 &  0.57 &  ...\\[0.5ex]
080714425 &  080714C &  $1.984\pm1.040$ &  $1.241\pm0.108$ &  $40.192\pm1.145$ &  -22.987  & 55.766 &   0.64 &  0.40 &  ...\\[0.5ex]
080714745 &  080714A &  $1.384\pm0.143$ &  $0.620\pm0.054$ &  $59.649\pm11.276$ &  -29.747  & 88.936 &   0.74 &  0.33 &  ...\\[0.5ex]
080715950 &  080715A &  $<0.101$ &  $0.011\pm0.001$ &  $7.872\pm0.272$ &  -2.203  & 11.867 &   0.90 &  0.20 &  ...\\[0.5ex]
080717543 &  080717A &  $0.568\pm0.194$ &  $0.369\pm0.032$ &  $36.609\pm2.985$ &  -23.414  & 49.420 &   0.53 &  0.34 &  ...\\[0.5ex]
080719529 &  080719A &  $1.990\pm0.830$ &  $1.241\pm0.108$ &  $16.128\pm17.887$ &  -12.339  & 19.739 &   0.64 &  0.40 &  ...\\[0.5ex]
080723557 &  080723B &  $0.040\pm0.017$ &  $0.027\pm0.002$ &  $58.369\pm1.985$ &  -0.180  & 89.694 &   0.22 &  0.15 &  ...\\[0.5ex]
080723913 &  080723C &  $0.013\pm0.004$ &  $0.008\pm0.001$ &  $0.192\pm0.345$ &  -0.160  & 0.224 &   0.96 &  0.58 &  ...\\
\end{longtable}
\tablecomments{
Redshift values marked with $\star$ are taken from \url{http://www.mpe.mpg.de/~jcg/grbgen.html}. 
 }
\end{scriptsize}
\end{center}

\end{document}